\documentclass[a4paper,11pt]{article}
\pdfoutput=1 

\usepackage{jcappub} 

\usepackage[T1]{fontenc} 
\usepackage[normalem]{ulem}

\def\hhref#1{\href{http://arxiv.org/abs/#1}{#1}} 

\title{\boldmath 
Searches for gamma-ray lines and `pure WIMP' spectra from Dark Matter annihilations in dwarf galaxies with H.E.S.S.}

\author{H.E.S.S. Collaboration, }
\author[1]{H.~Abdalla, }
\author[3,4,5]{F.~Aharonian, }
\author[3]{F.~Ait~Benkhali, }
\author[19]{E.O.~Ang\"uner, }
\author[39]{M.~Arakawa, }
\author[1]{C.~Arcaro, }
\author[23]{C.~Armand, }
\author[14]{M.~Arrieta, }
\author[8,1]{M.~Backes, }
% ...

\affiliation[1]{Centre for Space Research, North-West University, Potchefstroom 2520, South Africa}  
% ...
\affiliation[3]{Max-Planck-Institut f\"ur Kernphysik, P.O. Box 103980, D 69029 Heidelberg, Germany}  
\affiliation[4]{Dublin Institute for Advanced Studies, 31 Fitzwilliam Place, Dublin 2, Ireland}  
\affiliation[5]{National Academy of Sciences of the Republic of Armenia,  Marshall Baghramian Avenue, 24, 0019 Yerevan, Republic of Armenia } 
% ...
\affiliation[8]{University of Namibia, Department of Physics, Private Bag 13301, Windhoek, Namibia} 
% ...
\affiliation[14]{LUTH, Observatoire de Paris, PSL Research University, CNRS, Universit\'e Paris Diderot, 5 Place Jules Janssen, 92190 Meudon, France} 
% ...
\affiliation[19]{Aix Marseille Universit\'e, CNRS/IN2P3, CPPM, Marseille, France} 
% ...
\affiliation[23]{Laboratoire d'Annecy de Physique des Particules, Univ. Grenoble Alpes, Univ. Savoie Mont Blanc, CNRS, LAPP, 74000 Annecy, France} 
% ...
\affiliation[39]{Department of Physics, Rikkyo University, 3-34-1 Nishi-Ikebukuro, Toshima-ku, Tokyo 171-8501, Japan} 
% ...

\emailAdd{contact.hess@hess-experiment.eu}

\abstract{Dwarf spheroidal galaxies are among the most promising targets for detecting signals of Dark Matter (DM) annihilations. 
The H.E.S.S. experiment has observed five of these systems for a total of about 130 hours.  The data are re-analyzed here, and, in the absence of any detected signals, are  interpreted  in terms of limits on the DM annihilation cross section. Two scenarios are considered: i) DM annihilation into mono-energetic gamma-rays and ii) DM in the form of pure WIMP multiplets that, annihilating into all electroweak bosons, produce a distinctive gamma-ray spectral shape with a high-energy peak at the DM mass and a lower-energy continuum. 
For case i), upper limits at 95\% confidence level of about $\langle \sigma v \rangle \lesssim 3 \times 10^{-25}$ cm$^3$ s$^{-1}$ are obtained in the mass range of 400 GeV to 1 TeV. For case ii), the full spectral shape of the models is used and  several excluded regions are identified, but the thermal masses of the candidates are not robustly ruled out.}

\begin{document}
\maketitle
\flushbottom

\section{Introduction}
\label{sec:intro}


Cosmology and astrophysics deliver convincing evidence of the presence of Dark Matter (DM) in the Universe, and in particular in our Galaxy and its satellites~\cite{Bergstrom:2000pn,Bertone:2004pz}. However, its fundamental nature is still unknown. 
The hypothesis of Dark Matter being a new elementary particle with a typical mass around a TeV has gained a prominent status in the past decades, mostly because many New Physics theories, such as SuperSymmetry (SuSy), predict the existence of new states at this scale~\cite{Jungman:1995df}. While this is still arguably a theoretically appealing scenario, the empty-handed searches at the Large Hadron Collider at CERN~\cite{Aaboud:2018jiw} have started to push some of the theories into uncomfortable corners of their parameter space (in particular: towards large masses). The same searches also cast doubt on the very foundational principle of the popularity of these models, i.e. the belief that New Physics just above the weak scale is needed to protect the Higgs mass parameter from large quantum corrections (see e.g.~\cite{Martin:1997ns}).

Nevertheless, the TeV scale remains very attractive due to the so-called `WIMP (\textit{Weakly Interacting Massive Particles}) miracle': the realization that a particle with `weak mass' (i.e. of the order of the hundreds of GeVs to TeVs) and weak interactions can naturally produce the relic abundance of DM observed in cosmology via thermal freeze-out~\cite{Bertone:2004pz}. Such a particle retains these good (`miraculous') qualities of a WIMP independently of other possible theoretical considerations.


It is therefore as timely and relevant as ever to pursue searches for DM signals that are independent of any specific theory constructions such as SuSy but that point to TeV and multi-TeV masses. 
Arguably the most model-independent search is the one for monochromatic lines: these arise at $E \simeq M_{\rm DM}$ whenever DM annihilates directly into a pair of particles much lighter than the DM itself and including at least one photon. 
Another paradigmatic signal in this respect originates from `pure WIMP' models, further discussed in Sec.~\ref{sec:limitspureWIMPs}, in which DM annihilates both into $\gamma\gamma$ pairs and into gauge bosons, producing a distinctive gamma-ray spectral shape with a high-energy peak and a lower-energy continuum.

\medskip

\medskip

The High Energy Stereoscopic System, H.E.S.S.,~is in a good position to produce significant results in this context, as its natural { energy range} extends to the multi-TeV region. It has a very good energy resolution (in the relevant {energy range}) that allows the discrimination of sharp features in the spectra. 
A very significant amount of data has been accumulated with this instrument in the observation of promising targets for Dark Matter annihilation: previous H.E.S.S. publications have shown limits for continuum and line-like signals from the Galactic Centre (GC)~\cite{HESSlinesGC,Abdallah:2016ygi, Abdallah:2018qtu}, and for a continuum DM signal from dwarf { spheroidal} galaxies~\cite{Abramowski:2014tra}.

Dwarf spheroidal galaxies (dSphs) bound to the Milky Way (MW) are among the most promising targets for DM indirect detection. They are believed to be DM dominated~\cite{Courteau:2013cjm} and hence offer a large expected signal-to-noise ratio. In addition, they are close enough to provide a significant expected flux. Dwarf spheroidals are subject to astrophysical uncertainties (see e.g. Section 3.4 of~\cite{Lefranc:2016dgx})
that are independent of those of other promising targets, { like the GC,} making their observation crucial to establish a limit or a possible discovery of a signal.
For instance, their small baryonic content makes them less sensitive to the poor knowledge of the astrophysical processes involving baryons. { The related uncertainties} have an important impact on other targets that are baryon dominated, most notably the { GC},
which both H.E.S.S.~\cite{HESSlinesGC,Abdallah:2018qtu} and H.E.S.S.-II \cite{Abdalla:2016olq} have already observed
in the search for gamma-ray lines. Indeed, assuming a DM core in the inner few kiloparsecs of the MW weakens the limits on DM annihilation by orders of magnitude, with respect to the case of a peaked DM profile like Navarro-Frenk-White (NFW) or Einasto (see~\cite{HESS:2015cda}, and~\cite{Lefranc:2016fgn} for a recent quantification). The existence of such a core cannot be excluded by the results of DM simulations (which depend on the way baryonic effects are modelled, see e.g.~\cite{DiCintio:2013qxa,Marinacci:2013mha,Tollet:2015gqa,Mollitor:2014ara}),
and has even been argued to be favoured by observations~\cite{Nesti:2013uwa,Cole:2016gzv} (see however~\cite{Wegg:2016jxe} for another view). These considerations add motivation to search for gamma-ray like signals from dSphs.

As discussed in Sec.~\ref{sec:analysis}, H.E.S.S.~has observed 5 dSphs
(see e.g.~\cite{Abramowski:2014tra,Abramowski:2010aa,Aharonian:2007km} for previous related H.E.S.S. analyses). The scope of this paper is therefore to search for signals of Dark Matter annihilations in the dSphs observed by H.E.S.S., focussing on two generic spectral features: i) truly model-independent monochromatic gamma-ray lines and ii) peculiar shapes characterized by a high-energy peak and a lower-energy continuum that arise in the broad class of `pure WIMP' DM models, where by `pure WIMP' we mean multiplets of the SM weak group (see Sec.~\ref{sec:limitspureWIMPs}). {These searches are sensitive to the DM annihilation} cross section as a function of the DM mass, in the range 300 GeV to 20 TeV.

\bigskip

This paper is organized as follows. In Sec.~\ref{sec:analysis}  the features of the H.E.S.S.~instrument are briefly described and the observational procedure is discussed. In Sec.~\ref{sec:fluxlimits} the results are presented in terms of upper limits at 95\% confidence level (C.L.) to mono-energetic gamma-ray fluxes from the targets, without connection to DM. In Sec.~\ref{sec:DMlimits}  the corresponding constraints on the DM parameter space are derived: first the expected flux and then the bounds for the two cases { mentioned} above are presented. In Sec.~\ref{sec:discussion} the results are put into context, comparing them to {previous} constraints. In Sec.~\ref{sec:summary} the findings are summarized. 


\section{H.E.S.S. observations and analysis}
\label{sec:analysis}

\subsection{The H.E.S.S. instrument }
The H.E.S.S. experiment consists of an array of Imaging Atmospheric \v Cerenkov Telescopes (IACTs) located in the Khomas Highland of Namibia at an elevation of 1800 m. The original array of four {12} m diameter telescopes at the corners of a 120 m x 120 m square (H.E.S.S. I) began operation in 2003 and was enhanced by the addition of a {28} m diameter telescope at the centre of the grid in 2012 (H.E.S.S. II). The results presented in this paper make use of data collected by the initial H.E.S.S. I array.  

\subsection{Observations and data analysis}
{Observations of five dSphs, Fornax dSph (2010), Coma Berenices dSph (from 2010 to 2013), Sculptor dSph (2008), Carina dSph (from 2008 to 2010) and Sagittarius dSph (from 2006 to 2012), were carried out with the H.E.S.S. telescopes}. With the exception of the Fornax dSph, the data of all dSphs were collected in \textit{wobble mode}, i.e. the source was offset from the centre of the field of view in order to simultaneously measure the background flux and the flux from the region of interest. In the case of Fornax, observations were performed with large offsets since this galaxy was not the primary target of the telescopes. Only data meeting the standard quality control criteria for data acquisition and weather conditions were used in this analysis~\cite{Aharonian:2006pe}. {An additional quality cut of a minimum number of three telescopes in operation during the observation runs was applied in order to improve the gamma-ray events energy reconstruction and direction~\cite{Aharonian:2006pe}. The total data after quality cuts amounts to a livetime of 6.0 h for Fornax, 10.9 h for Coma Berenices, 11.8 h for Sculptor, 22.9 h for Carina, and 85.5 h for Sagittarius. The average zenith angle of these obervations are 14$^{\circ}$ for Fornax, 48$^{\circ}$ for Coma Berenices, 14$^{\circ}$ for Sculptor, 34$^{\circ}$ for Carina, and 16$^{\circ}$ for Sagittarius.

For this work the analysis was carried out using a \textit{model analysis}~\cite{model} chain with \textit{standard cuts} to select signal events and suppress background. In this method, the observed intensity is compared pixel by pixel with a pre-calculated semi-analytic shower model. The results of the analysis were fully compatible with a crosscheck using an independent calibration and analysis chain~\cite{ImPACT}. The gamma-ray signal was searched in a circular (ON) region with angular radius $\theta <$0.2$^{\circ}$ for Fornax, Sculptor, Carina and Sagittarius, and with an angular radius $\theta <$0.3$^{\circ}$ for Coma Berenices due to the expected extension of their DM halo in the sky~\cite{Geringer-Sameth:2014yza}.
{The background was determined by the reflected background technique~\cite{Aharonian:2006pe}, where several (OFF) regions with the same size and located at the same offset from the observation position as the ON-region are used for its measurement (see Fig. 9 of ref.~\cite{Aharonian:2006pe}). A signal excess is calculated using the number of events in the ON-region, N$_{\rm ON}$, number of events in the OFF-region, N$_{\rm OFF}$, and the exposure ratio between control and signal regions, $\alpha$. The excess significance is estimated using the Li\&Ma method~\cite{lima}. The results of the analysis for the five dSph galaxies are summarised in Table~\ref{tab1}}. Note that this same data set has already been analysed in a previous work~\cite{Abramowski:2014tra} with a different calibration, analysis chain and signal extraction region, which leads to slightly different livetimes and analysis results. All results from the three analyses are compatible with no gamma-ray signal excess.}

\begin{table}[!htb]
\small
\centering
\begin{tabular}{l|ccccccc}
\hline\hline
Source name &  livetime (h) & zenith & offset & N$_{\rm ON}$ & N$_{\rm OFF}$ & $\alpha$ & significance ($\sigma$) \\
\hline
Fornax & 6.0 & 14$^{\circ}$ &  1.7$^{\circ}$ &192 &4943 &21.6 &-2.4 \\ 
Coma Berenices &10.9 & 48$^{\circ}$ &  0.4$^{\circ}$  &1513 &4191 &2.8 &0.4 \\
Sculptor & 11.8 & 14$^{\circ}$ &  0.7$^{\circ}$  &1176 &11141 &8.9 &-2.0 \\
Carina &  22.9 & 34$^{\circ}$ &  0.7$^{\circ}$  &1515 &11865 &8.0 &0.9 \\
Sagittarius &  85.5 & 16$^{\circ}$ &  0.7$^{\circ}$  &8305 & 67715 &8.3 &1.2 \\
\hline\hline
\end{tabular}
\caption{ \label{tab1} Summary of data analysis results for each dSph: live time (not corrected for acceptance), average zenith angle of observation, average observation offset from the centre of the field of view, number of events in signal region (ON-region), number of events in control region (OFF-region), exposure ratio between control and signal regions $\alpha$, and excess significance~\cite{Aharonian:2006pe}. The differences between these numbers and those of ref.~\cite{Abramowski:2014tra} come mainly from the change in the analysis chain and signal extraction region.}\end{table}


\section{Flux upper limits to mono-energetic lines}
\label{sec:fluxlimits}

Since no excess was found in the signal extraction region ($\theta <$0.3$^{\circ}$ for Coma Berenices, $\theta <$0.2$^{\circ}$ for the other dSphs), flux upper limits (ULs) on a monochromatic line-like signal of energy $E_{\gamma}$ were calculated. 

A 2-dimensional (2D) likelihood function binned in energy and spatial coordinates is used to calculate limits on the gamma-ray flux, taking advantage of the spatial and spectral differences between the expected DM signal and measured residual cosmic-ray background. 

For the case at hand, the energy range is divided into 40 logarithmically-spaced bins between 100 GeV and 100 TeV. The spatial regions of interest (RoIs) are defined as circular concentric regions of 0.1$^{\circ}$ width each, and centered at the dynamical centre of the dSph galaxy. Due to the small angular extension of dSphs~\cite{Geringer-Sameth:2014yza}, for Sculptor, Carina, Fornax and Sagittarius, only two spatial bins (from 0$^{\circ}$ to 0.2$^{\circ}$ radius), and for Coma Berenices three spatial bins (from 0$^{\circ}$ to 0.3$^{\circ}$ radius) are used. 

The likelihood is calculated for each considered energy $E_{\gamma}$ as a product over the spatial RoIs (bins $i$) and the energy bins (bins $j$).  The 2D binned spatial and spectral likelihood formula is composed of a Poisson ``ON'' term (first term in Eq.~(\ref{eqn1})) and a Poisson ``OFF'' term (second term in Eq.~(\ref{eqn1})):
\begin{equation}
\mathcal{L}_{ij}(s_{ij},b_{ij}| \mathcal{D}_{ij} ) = {\frac{(s_{ij}+b_{ij})^{N_{{\rm ON},ij}}}{N_{{\rm ON},ij}!}}e^{-(s_{ij}+b_{ij})}\times \frac{(\alpha_{i}b_{ij})^{N_{{\rm OFF},i,j}}}{N_{{\rm OFF},ij}!}e^{-(\alpha_{i}b_{ij})}
\label{eqn1}
\end{equation}
where the $\mathcal{D}_{ij}$ is the set containing the number of gamma-ray events observed in the signal region ($N_{{\rm ON},ij}$), in the control region ($N_{{\rm OFF},ij}$), and exposure ratio between control and signal regions $\alpha_i$, as defined in Table~\ref{tab1}, and calculated in each bin ($i,j$). $s_{ij}$ is the number of predicted signal events in the bin ($i,j$) and $b_{ij}$ the number of expected background events. The predicted signal is computed by folding the considered gamma-ray flux with the H.E.S.S. response functions, for the specific observation conditions of the analyzed dataset. The expected signal in a spatial bin $i$ and energy bin E$_{j}$ is given by

\begin{equation}
s_{ij} = T_{\rm obs} \int_{\Delta E_j} dE_{\gamma}^r \int_0^{\infty} dE_{\gamma}^t \frac{{\rm d}\Phi_{\gamma,i}(E_{\gamma}^t)}{{\rm d} E_{\gamma}^t} \times A_{\rm eff}(E_{\gamma}^t) \times {\rm PDF}(E_{\gamma}^t,E_{\gamma}^r)
\end{equation}   
where $T_{\rm obs}$ is the observation time, $E^t_{\gamma}$ is the true energy, $A_{\rm eff}$ is the effective collection area as function of the true energy, and ${\rm PDF}(E_{\gamma}^t,E_{\gamma}^r)$ is the representation of the energy resolution as the probability density function $P(E_{\gamma}^r|E_{\gamma}^t)$, of observing an event at the reconstructed energy $E_{\gamma}^r$ for a given true energy $E_{\gamma}^t$.  The predicted differential flux of a mono-chromatic line of energy $E_{\gamma}$ is given in this case by ${\rm d}\Phi_{\gamma,i}(E_{\gamma}^t) / {\rm d} E_{\gamma}^t =  \Phi_{0,i} \delta(E_{\gamma}^t - E_{\gamma})$, where $\delta(E)$ is a Dirac delta function, and $\Phi_{0,i}$ is the normalization in each spatial bin, which is considered constant for the flux upper-limits calculation. The convolution of the H.E.S.S. resolution with the line-like signal results in an expected signal with a log-normal shape with an r.m.s. $\sigma_E/E$ varying between 10\% and 20\% for energies above 200 GeV~\cite{model}. The exact shape of the energy resolution depends on the gamma-ray energy and on the parameters of the observation data sets, e.g. the mean zenith angle, the optical efficiency, the mean off-axis angle and the cuts used in the analysis~\cite{Aharonian:2006pe, model}. 

The total likelihood function over the full energy range and all the spatial RoIs is the product of the individual likelihood functions over all bins $i$ and $j$,

\begin{equation}
\mathcal{L} (s,b|\mathcal{D}) = \prod_{ij} \mathcal{L}_{ij}(s_{ij},\hat{b}_{ij}| \mathcal{D}_{ij} ) \, ,
\end{equation}
where $\hat{b}_{ij}$ correspond to the best estimate of the background for a given signal $s$, found by solving ${\rm d} \mathcal{L}/{\rm d}b_{ij}  = 0$. The likelihood function can be written as function of the integrated flux $\Phi_{\gamma}$ in the whole energy range. Constraints on the integrated flux are obtained with the likelihood ratio test statistics given by
\begin{equation}
{\rm TS} = -2\ {\rm ln}\left(\frac{\mathcal{L} (\Phi_{\gamma},\hat{\hat{b}})}{\mathcal{L} (\hat{\Phi}_{\gamma},\hat{b})} \right)
\end{equation}
where $\hat{\hat{b}}$ is the value of background that maximizes the likelihood function $\mathcal{L}$ at $\Phi_{\gamma}$, and $\hat{\Phi}_{\gamma}$ is the value of flux that globally maximizes $\mathcal{L}$. In case $\hat{\Phi}_{\gamma}$ is found to be negative, its value is shifted to zero, as negative fluxes are not physical. The TS function follows an approximate $\chi^2$ distribution, and values $\Phi_{\gamma}$ for which TS is higher than 2.71 are excluded at 95\% C.L.. The {\sc Minuit2} subroutine {\sc Migrad} ~\cite{minuit} is used for the minimization and confidence interval computation. 

\begin{figure}[t]
\centering
\includegraphics[width=0.7\textwidth]{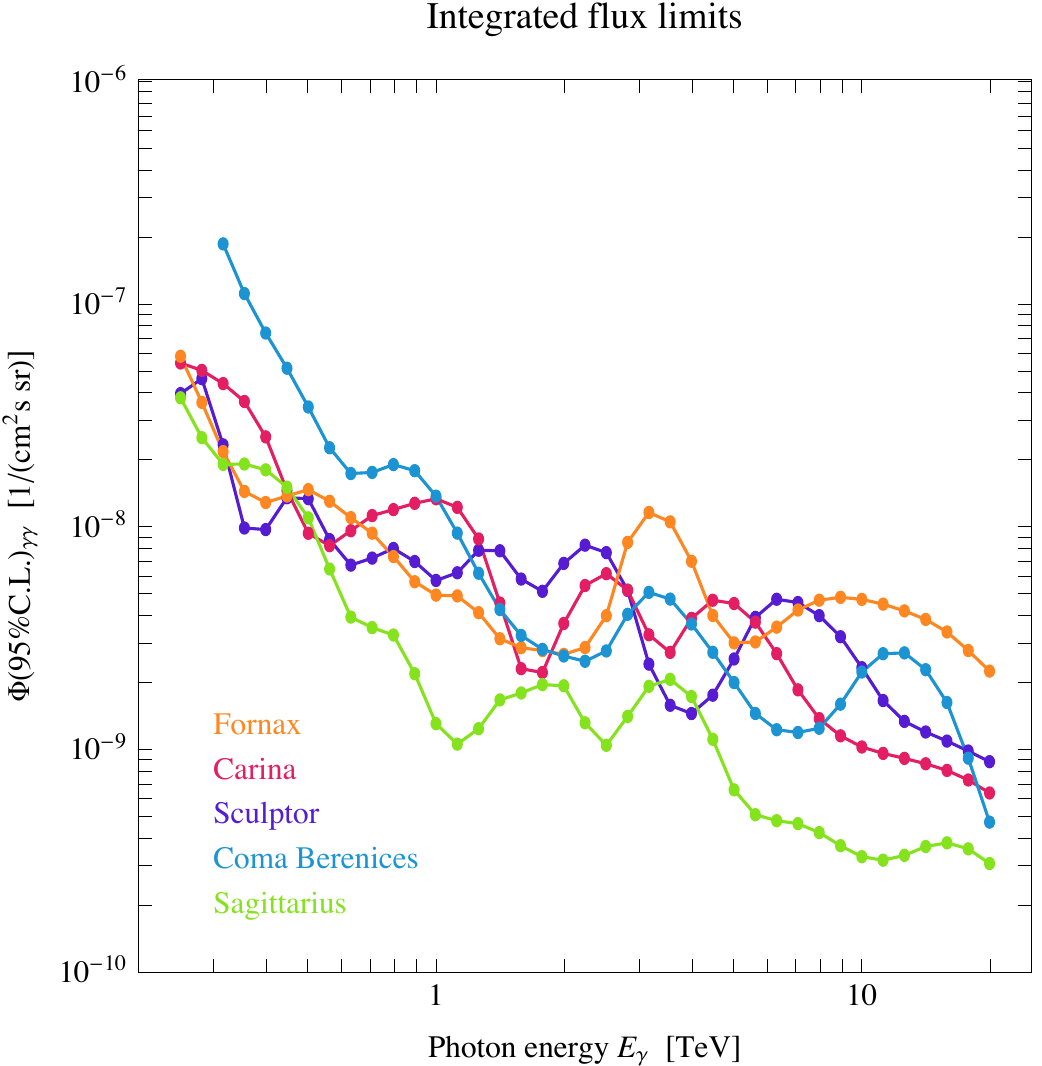} 
\caption{\label{fig:fluxlimits} Upper limits at 95\% C.L. on the integrated flux as a function of photon energy $E_\gamma$ for Fornax, Carina, Coma Berenices, Sculptor and Sagittarius dSphs. The dots correspond to the discrete values of photon energy which are actually tested.} 
\end{figure}

The calculated ULs on the integrated flux at the 95\% C.L. are shown in Figure~\ref{fig:fluxlimits} as a function of $E_{\gamma}$. The differences in flux limits from the five dSphs arise from {statistical fluctuations}, variations in observation conditions and observation time for each galaxy. For instance, the large zenith angles of observation of Coma Berenices dSph causes a loss of effective area at low energies, and hence the upturn of its limits below $\sim$700 GeV.



\section{Exclusion limits on the Dark Matter annihilation cross section}
\label{sec:DMlimits}

\subsection{Analysis methodology}
\label{sec:methodology}

In general terms, the gamma-ray flux from the annihilations of (self-conjugated) DM particles of mass $M_{\rm DM}$ in a DM halo is given by a particle physics term times an astrophysics term:\\

\begin{equation}
\label{eq:diff_flux}
\frac{{\rm d}\Phi_{\gamma,i}(\Delta\Omega_i,E_{\gamma})}{{\rm d} E_{\gamma}}\,= \frac12 \frac{1}{4\pi}\,\underbrace{\sum_k \frac{\langle \sigma v \rangle_k}{M_{\rm DM}^2}
	\frac{{\rm d} N^k}{{\rm d}E_\gamma}}_{\rm Particle\,
	Physics} \,\times\,\underbrace{J(\Delta\Omega_i)}_{\rm Astrophysics} \, ,
\end{equation}
where the astrophysical factor, also called $J$-factor, is calculated inside each spatial bin $i$ considered here, and it is defined as\\
\begin{equation}
J(\Delta\Omega_i) = \int_{\Delta \Omega_i}  \int_{\rm l.o.s.} {\rm d}\Omega \, {\rm d} s \ \rho_{\rm DM}^2[r(s,\Omega)] .
\label{eq:Jfactors}
\end{equation}

In equation~(\ref{eq:Jfactors}) the squared density distribution of DM ($\rho_{\rm DM}^2$) is integrated along the line of sight (l.o.s.)~and over $\Delta\Omega_i$, the solid angle of each spatial bin. The coordinate $r$ is defined by $r^2 = D^2 + s^2 - 2\, D\, s\, {\rm cos}\,\theta$, where $s$ is the distance along the line of sight, $\theta$ is the angle between the observation direction and the centre of the dSphs, and $D$ is the distance of the dSph. The particle physics term contains the DM particle mass, $M_{\rm DM}$, the velocity-weighted annihilation cross section, $\langle \sigma v\rangle$, and the differential spectrum of gamma rays from all final states weighted by their corresponding branching ratios, ${\rm d} N^i/{\rm d} E_{\gamma}$.

\bigskip




The $J$-factor in eq.~(\ref{eq:Jfactors}) is determined via dynamical mass models that are based on the Jeans equation.
For each galaxy, such models are fit to the kinematical data of those stars that are { classified as} good `tracers' of the gravitational potential of the galaxy.
For any given model, the statistical uncertainty of the resulting $J$-factor { is typically small} for classical dSphs, that contain a lot of stellar tracers.
However, both a different modelling and a different set of { astrophysical} assumptions 
might give much different results for the $J$-factors, inducing systematic errors which are not evident from a single study (see e.g. Section 3.4 of~\cite{Lefranc:2016dgx} for a thorough discussion).
The past few years have seen an intense activity in the determination of the above $J$-factors (see e.g.~\cite{Martinez:2013els, Geringer-Sameth:2014yza, Bonnivard:2015xpq,Hayashi:2016kcy, Ullio:2016kvy, Evans:2016xwx, Genina:2016kzg}), that has allowed us to identify some { dSphs} that are more robust against the above systematics.

In this work, for Fornax, Coma Berenices, Sculptor and Carina, the results of ref.~\cite{Geringer-Sameth:2014yza} are used.
In the case of Sagittarius the results from ref.~\cite{Viana:2012zz} are used, where a detailed modeling of the DM halo was obtained in the case of an isothermal DM profile, using an evolutionary N-body simulation of the dSph to describe the observed structural and kinematical distributions of stars in the tidal tails within the MW. A truncation of the DM halo at a radius of 4 kpc is introduced to account for the tidal disruption observed in this system. The $J$-factors that are adopted, as well as their uncertainties, are reported in Table~\ref{tab:Jfactors}. 

Constraints on the velocity-weighted annihilation cross section $\langle\sigma v\rangle$ are obtained following the same procedure described in section~\ref{sec:fluxlimits}. Here the likelihood function of each galaxy is written as function of $\langle\sigma v\rangle$ and the J-factor, \emph{i.e.} $\mathcal{L}^{\rm dSph} (\langle \sigma v \rangle,J, b)$. In order to account for the uncertainties in the $J$-factors, a nuisance parameter is added to the joint likelihood function. For each galaxy a log-normal distribution of $J$ is convolved with the likelihood function
\begin{equation}
\mathcal{\tilde{L}}^{\rm dSph} (\langle \sigma v \rangle,J, b) = \mathcal{L}^{\rm dSph} (s,b|\mathcal{D}) \times \frac{1}{{\rm ln}(10) J \sqrt{2\pi}\sigma_J} e^{-\left[ {\rm log}_{10}(J) - \overline{{\rm log}_{10}J} \right]^2 / 2\sigma^2_J}
\end{equation}   
where $\overline{{\rm log}_{10}J}$ and $\sigma_J$ are the mean and standard deviations
of the distribution of ${\rm log}_{10}(J)$, respectively, and their values are given in Table~\ref{tab:Jfactors}. 

\begin{table}[!htb]
	\small
	\centering
	\begin{tabular}{l|ccc|ccc}
		\hline\hline
		Source name & distance & \multicolumn{2}{c|}{position}  &$\overline{{\rm log}_{10}J}$ &$\sigma_J$ & ref. \\
		 & (kpc) & RA (h m s) & Dec ($^{\circ}\, ^\prime \, ^{\prime \prime}$) & ${\rm log}_{10} {\rm (GeV^{2} cm^{-5})}$ & ${\rm log}_{10} {\rm (GeV^{2} cm^{-5})}$ &  \\
		\hline 
		Fornax & 140 & 2 39 59.3 & -34 26 57 & 17.72 ($\theta=0.2^{\circ}$) & 0.18 & \cite{Geringer-Sameth:2014yza} \\
		Coma Berenices & 44 & 12 26 59 & 23 54 15 & 19.52 ($\theta=0.3^{\circ}$) & 0.37 & \cite{Geringer-Sameth:2014yza} \\
		Sculptor & 79 & 01 00 09 & -33 42 32 & 18.36 ($\theta=0.2^{\circ}$) & 0.12 & \cite{Geringer-Sameth:2014yza} \\
		Carina & 101 & 06 41 36 & -50 57 58 & 17.86 ($\theta=0.2^{\circ}$) & 0.10 & \cite{Geringer-Sameth:2014yza} \\
		Sagittarius & 25 & 18 55 03 & -30 28 42 & 18.34 ($\theta=0.2^{\circ}$) & 0.30 & \cite{Viana:2012zz} \\
		\hline\hline
	\end{tabular}
	\caption{Astrophysical properties of the dSph: distance in kpc, center position in RA-Dec coordinates,  $J$-factors calculated in the RoIs with total opening angle given in brackets, and 1$\sigma$ uncertainty extracted from the references in last column.  \label{tab:Jfactors} }
\end{table}

The total combined likelihood function of all the dSphs is simply the product of the individual likelihoods, 

\begin{equation}
\mathcal{\tilde{L}}^{\rm Comb} (\langle \sigma v \rangle) = \prod_{\rm dSph} \mathcal{\tilde{L}}^{\rm dSph} (\langle \sigma v \rangle,\hat{J}, \hat{b}) \, ,
\end{equation}
where $\hat{J}$ are the values of $J$-factor that maximize the individual likelihood functions $\mathcal{\tilde{L}}^{\rm dSph}$.\footnote{Note that we do not perform a strict joint-likelihood maximization because the systematical uncertainties of each dwarf (given by the nuisance parameter in the $J$-factor) are different, so we maximize them individually before we combine the likelihoods.}

\subsection{Limits on generic DM annihilation into mono-energetic gamma rays}
\label{sec:limitslines}

\begin{figure}[t]
\centering
\includegraphics[width=0.7\textwidth]{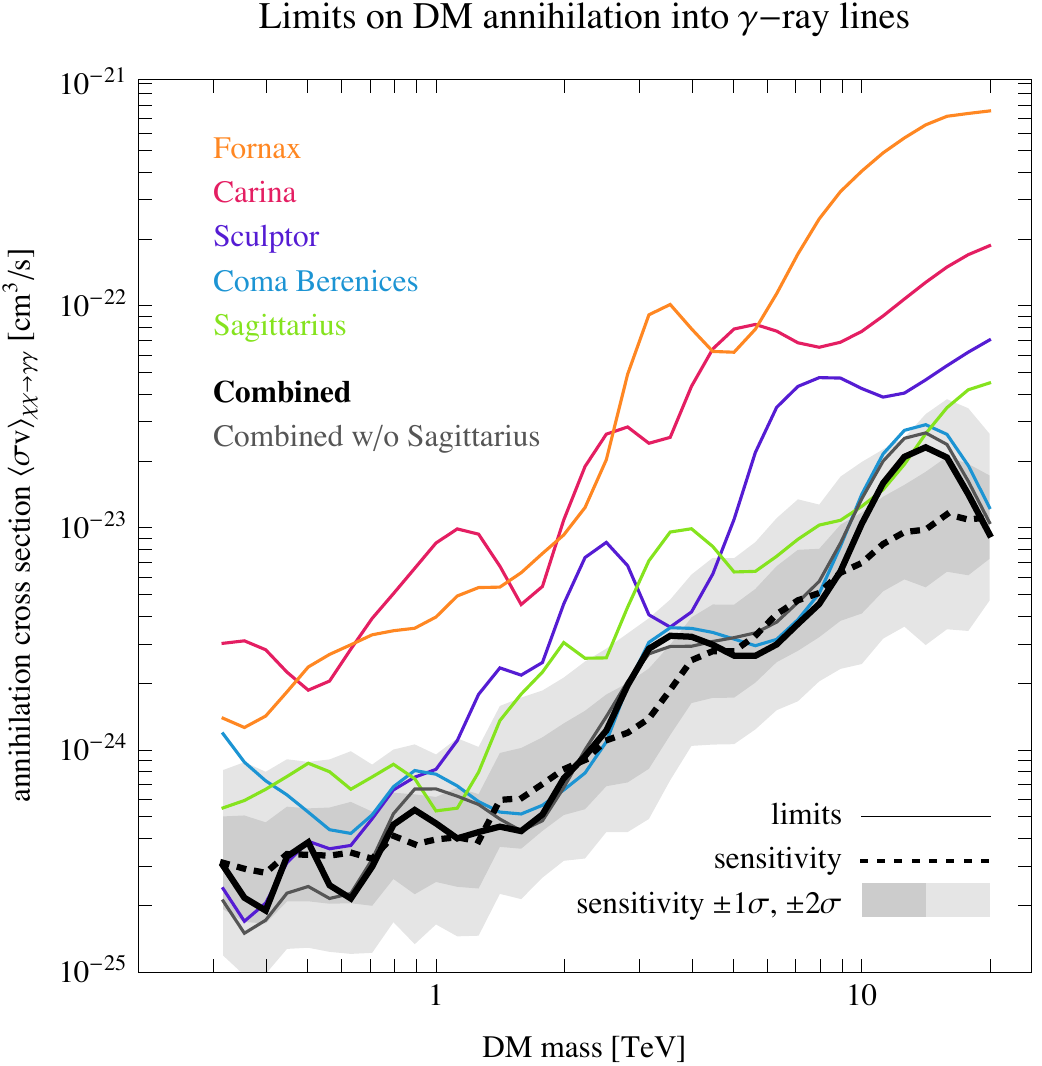} 
\caption{\label{fig:linelimits} 95\% C.L. exclusion limits on the velocity weighted cross section for DM self-annihilation into $\gamma\gamma$ as a function of $M_{\rm DM}$ for Fornax, Carina, Coma Berenices, Sculptor and Sagittarius dSphs.}
\end{figure}

Figure~\ref{fig:linelimits} shows the 95\% C.L. upper-limits on $\langle \sigma v \rangle$ versus $M_{\rm DM}$ in the case of a DM particle annihilating into $\gamma\gamma$ and producing a monochromatic line-like signal with energy $E_{\gamma} = M_{\rm DM} $. Limits for the individual dSphs as well as a combined result for the five galaxies, and for the four galaxies, excluding Sagittarius\footnote{Evidence exists that the Sagittarius galaxy is in a state of tidal disruption~\cite{Ibata:2000ys}, that could potentially invalidate the analysis that leads to the estimate of its J-factor~\cite{Viana:2012zz}. However, as seen in Figure~\ref{fig:linelimits}, the inclusion of Sagittarius has an impact of less than 50\% on the combined limits.}, are shown. 
A sensitivity curve of the combined limits, assuming a background-only hypothesis (i.e. $N_{{\rm ON},ij} = N_{{\rm OFF},ij}/\alpha_i$), is obtained from 500 Poisson realizations of both $N_{{\rm ON},ij}$ and $N_{{\rm OFF},ij}$. The mean sensitivity as well the statistical 68\% ($\pm$1$\sigma$) and 95\% ($\pm$2$\sigma$) containment bands are also plotted.

A limit of $\langle \sigma v \rangle \lesssim 3 \times 10^{-25}$ cm$^3$ s$^{-1}$ is reached in the mass range of 400 GeV to 1 TeV. The combination of all { five} galaxies allows an improvement in the constraints up to a factor of 2 around 600 GeV with respect to individual galaxies. 

Note that, at certain DM masses, the combined limit becomes worse than some individual limits, and the combined limit without Sagittarius becomes more constraining than the combined one that includes Sagittarius. This is due to the statistical effect of adding an individual data set with large negative fluctuations (or excesses) and large expected signal around those energies (case of Sculptor at $\sim$ 350 GeV, Coma Berenices at $\sim$ 2 TeV, and Sagittarius at $\sim$10 TeV), to large data sets with positive fluctuations and smaller or similar expected signal (case of Carina and Sagittarius at $\sim$ 350 GeV, Carina, Sculptor and Sagittarius at $\sim$ 2 TeV, and Coma Berenices at $\sim$10 TeV). The limits of the individual data sets will be highly overestimated, while the combination with the large data sets will push the combined limits to less constraining values.

\subsection{Limits on pure WIMP models}
\label{sec:limitspureWIMPs}

In this section `pure WIMP' models are briefly introduced and the distinctive shapes of their gamma-ray annihilation spectra are discussed. The results for the limits on this class of models are then presented.

\medskip

Pure WIMP models (or Minimal Dark Matter models~\cite{Cirelli:2005uq,Cirelli:2007xd,Cirelli:2008id,Cirelli:2009uv,DelNobile:2015bqo}) aim at introducing on top of the Standard Model the minimal amount of New Physics which is necessary to explain the DM problem. This means a new multiplet of particles $\chi$, charged under $SU(2)_L$. 
One of the components of this multiplet, the electrically neutral one~\footnote{Or one with a small millicharge~\cite{DelNobile:2015bqo}.}, constitutes the Dark Matter.
The charged partners are slightly heavier and decay very promptly into the neutral ones, so that they do not play any relevant cosmological role of their own. Due to their minimality, these models feature a phenomenology which is predictive and precisely computable: typically, for a given DM mass, the annihilation cross section (hence the expected gamma ray fluxes) is precisely determined. 

In this work, the focus is on two specific candidates (both with hypercharge = 0): the fermionic triplet ($\chi \equiv (\chi^+,\chi_0,\chi^-)$) and the fermionic quintuplet  ($\chi \equiv (\chi^{++},\chi^+, \chi_0,\chi^-,\chi^{--})$). 
The former requires an imposed symmetry to guarantee its stability. It turns out to have the same quantum numbers of pure wino DM in SuSy and is therefore interesting as a paradigm for a broader set of constructions (see also \cite{Cirelli:2014dsa}). 
The latter draws its appeal from the fact that it is indeed minimal, since its stability is guaranteed by a residual accidental symmetry, which is a consequence of the gauge and matter content of the SM (analogously to the baryon number, which makes the proton stable). It was first put forward in~\cite{Cirelli:2005uq}.

\begin{figure}[t]
\centering
\includegraphics[width=0.3\textwidth]{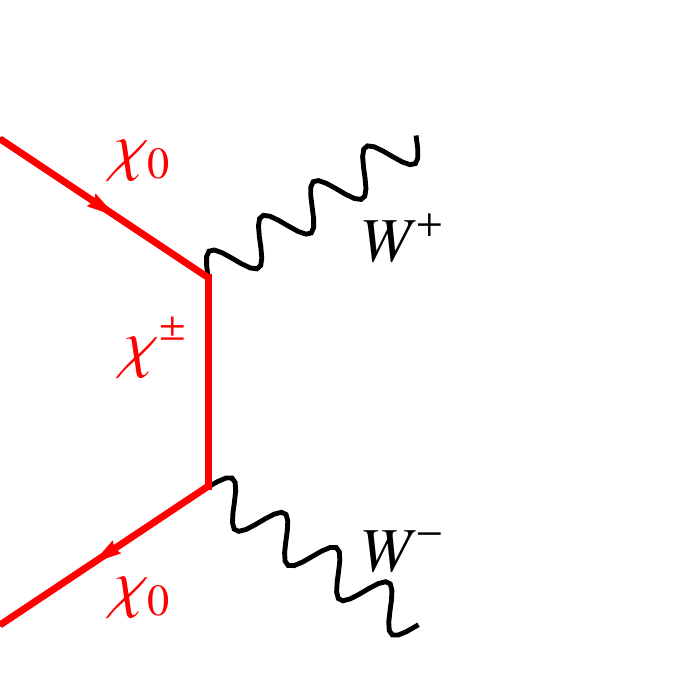} 
\includegraphics[width=0.3\textwidth]{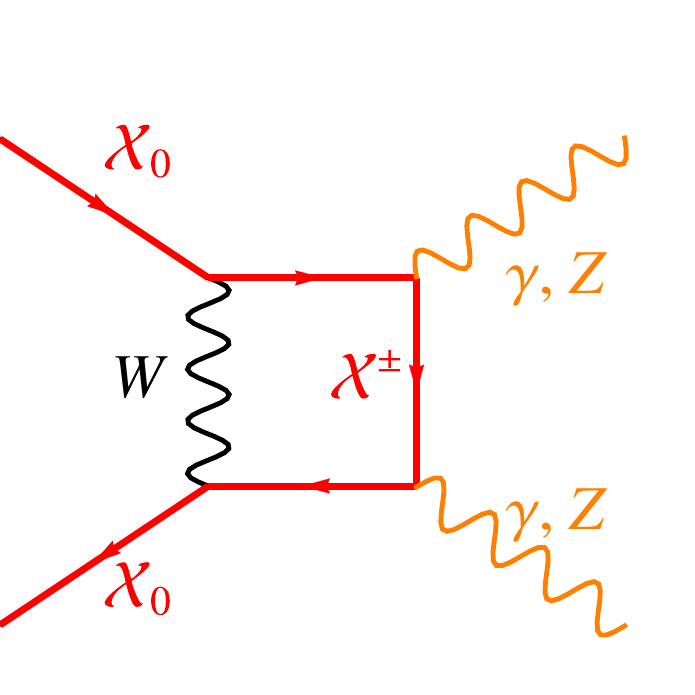} 
\caption{\label{fig:annihilation_channels}Illustration of the tree-level diagram and a 1-loop diagram relevant for annihilation signals in `pure WIMP' models. }
\end{figure}

By its very nature as a weakly interacting particle, the DM of these models features as the most relevant annihilation channel at the tree level into weak gauge bosons: $W^+W^-$.\footnote{Tree level annihilations into $ZZ$ are forbidden, in the cases that are considered here, by the $Y=0$ requirement.} At higher orders, the loop can be closed and therefore inevitably the annihilation into $\gamma\gamma$, $\gamma Z$ and $ZZ$ arises. 
Some of these annihilation diagrams are represented in Fig~\ref{fig:annihilation_channels}.  
The $W^+W^-$ and $ZZ$ channels produce a continuum gamma ray spectrum.\footnote{Except
for the case in which hard photons are radiated from a $W^\pm$, in which case a peak close in energy to the value of the DM mass arises. This is taken into account in the computations below by using the spectra as provided by {\sc Pppc4dmid}~\cite{Cirelli:2010xx}, that include the leading log-enhanced radiation, which is a good approximation for the masses and energy resolutions of our interest (see e.g.~\cite{Baumgart:2017nsr} for recent refinements).}
The $\gamma\gamma$ and $\gamma Z$ annihilation channels produce instead a gamma-ray line, and a much smaller continuum at lower energies. Therefore, in general terms, the gamma-ray spectrum of pure WIMP models features: a prominent line at $E_\gamma \simeq M_{\rm DM}$ and a continuum shoulder at $E_\gamma < M_{\rm DM}$ with the characteristic shape of gamma rays from electroweak (EW) showering and $W$ and $Z$ decays (see e.g.~\cite{Cirelli:2010xx}).

\medskip

\begin{figure}[!t]
\centering
\includegraphics[width=0.48\textwidth]{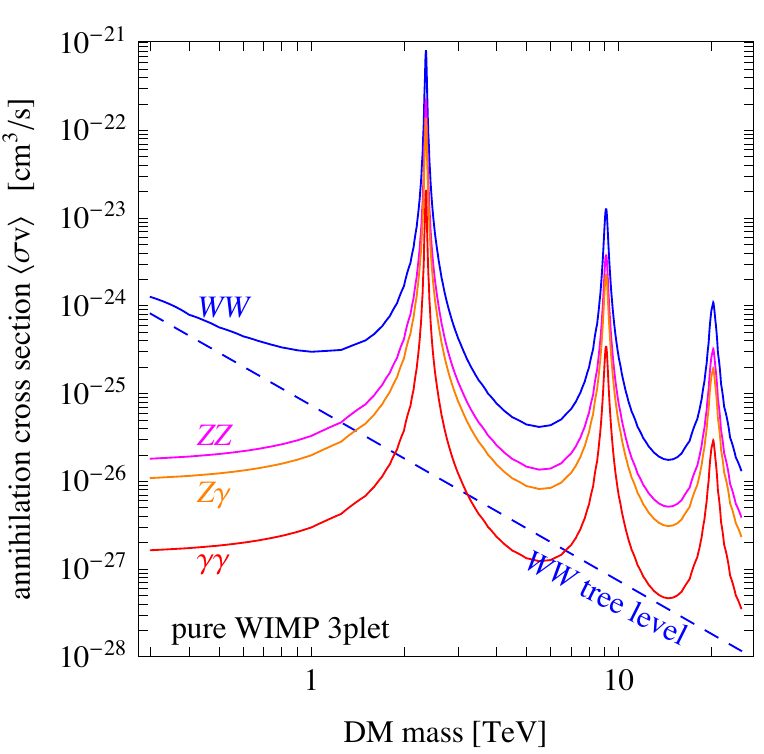}  \quad
\includegraphics[width=0.48\textwidth]{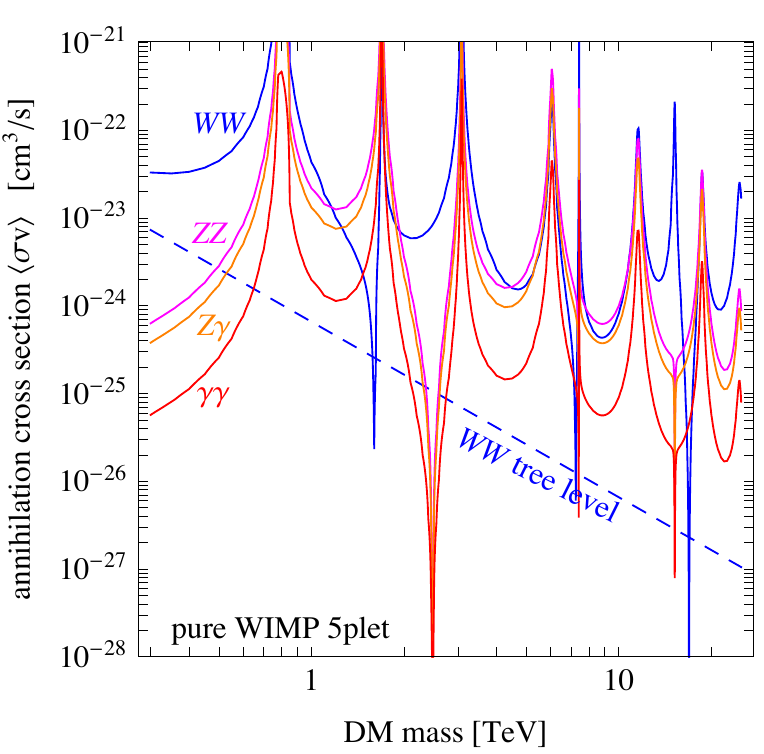} 
\caption{\label{fig:cross_sections} Annihilation cross sections of the pure WIMP 3plet (left) and the pure WIMP 5plet (right), including the Sommerfeld enhancement.}
\end{figure}

The ratio of the peak (from $\chi_0\chi_0 \to \gamma\gamma$) with respect to the continuum (from $\chi_0\chi_0 \to W^+W^-$) depends on the respective annihilation cross sections. While it is generally assumed that the peak, being generated at loop level, is suppressed with respect to the continuum, this is much less true in this class of models. This 
non-trivial result is a consequence of the so-called Sommerfeld enhancement. The importance of this non-perturbative phenomenon has been discussed, for multi-TeV DM models, extensively in the past (see~\cite{sommerfeld} for a non-exhaustive list). It arises in the presence of an effective long-range force among the DM particles in the annihilation process. In the case at hand, the interaction is just the EW one, mediated by the $W$ and $Z$ boson. Since the masses { of the EW gauge bosons} are much smaller than that of multi-TeV DM, the EW interaction becomes effectively long-range.
This is a non-perturbative effect, which has the power to:
i) enhance the annihilation cross sections by orders of magnitude with respect to the processes that do not include it, especially for some specific `resonant' values of the DM mass;
ii) increase the cross section for DM annihilation into $\gamma \gamma$ to values closer to that for the annihilation into $WW$.
As a concrete example, in Fig~\ref{fig:cross_sections} the different cross sections are shown, including the Sommerfeld enhancement, for the two models under consideration.\footnote{
In passing it should be noted that the formation of bound states can alter the precise position and amplitude of the peaks in the quintuplet case~\cite{Mitridate:2017izz}. 
The value of the thermal mass shown in Fig~\ref{fig:limitsmultiplet} (right) is corrected by this effect~\cite{Mitridate:2017izz}. Bound state formation is instead expected to be less important at present times in dSph's because, with respect to freeze-out time, the DM relative velocities are much smaller and only the $\chi_0 \chi_0$ initial state is available. In light of this observation, and in absence of quantitative calculations, this effect is not taken into account in the computation of the indirect signals. 
Bound state formation has instead been shown to have a negligible effect in the triplet case~\cite{Asadi:2016ybp} for the values of DM mass considered here.
}

\medskip

\begin{figure}[!t]
\centering
\includegraphics[width=0.48\textwidth]{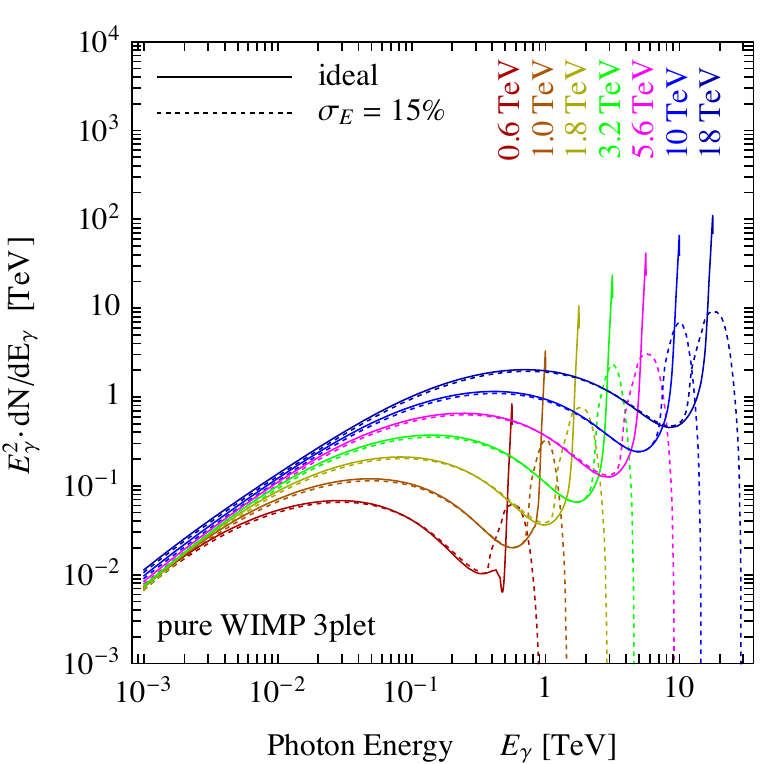} \quad
\includegraphics[width=0.48\textwidth]{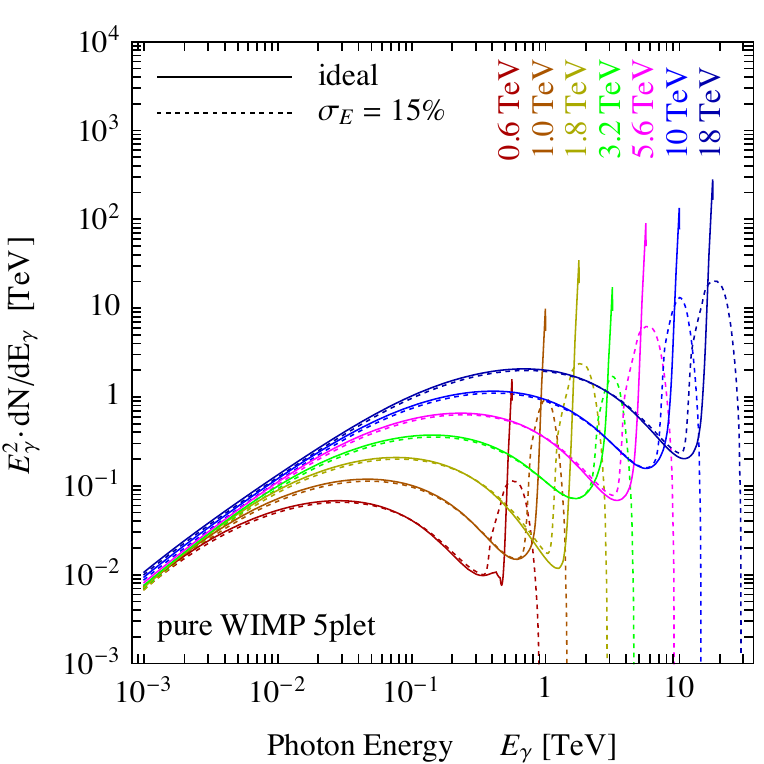} 
\caption{\label{fig:spectra} Gamma-ray spectra from pure WIMP 3plet (left) and pure WIMP 5plet (right) annihilations. Different masses have been considered and, for illustration, plotted with or without an indicative energy smearing of 15\% due to finite energy resolution.}
\end{figure}

Folding the annihilation cross section, for any given mass, with the gamma-ray yields (that are derived from {\sc Pppc4dmid}~\cite{Cirelli:2010xx} including EW corrections~\cite{Ciafaloni:2010ti}), allows the gamma-ray spectra to be univocally computed, shown in Fig~\ref{fig:spectra}. For illustration, the spectra smeared by the energy resolution function of the instrument are also plotted, as discussed in Sec.~\ref{sec:analysis}.

\begin{figure}[!t]
\centering
\includegraphics[width=0.48\textwidth]{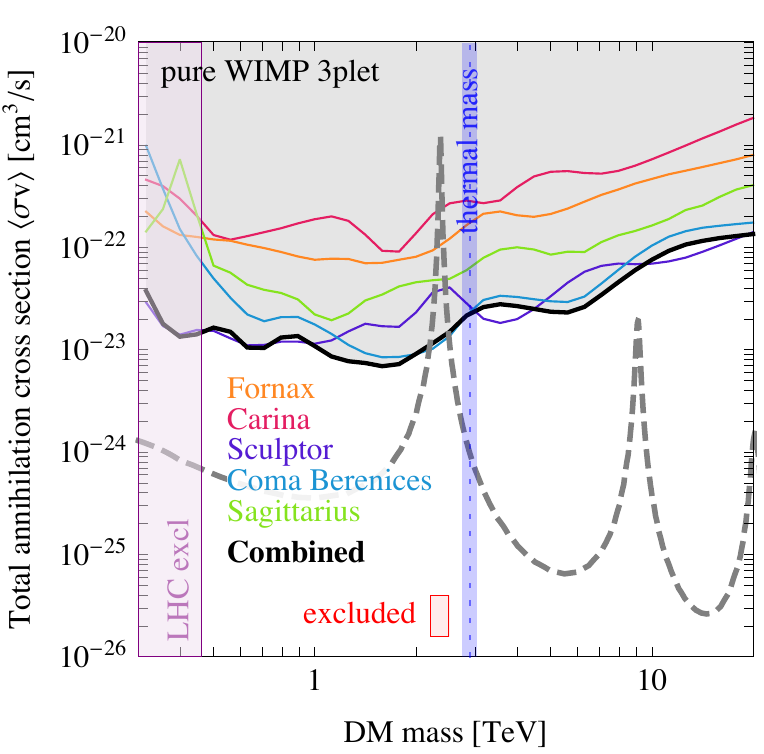} \quad
\includegraphics[width=0.48\textwidth]{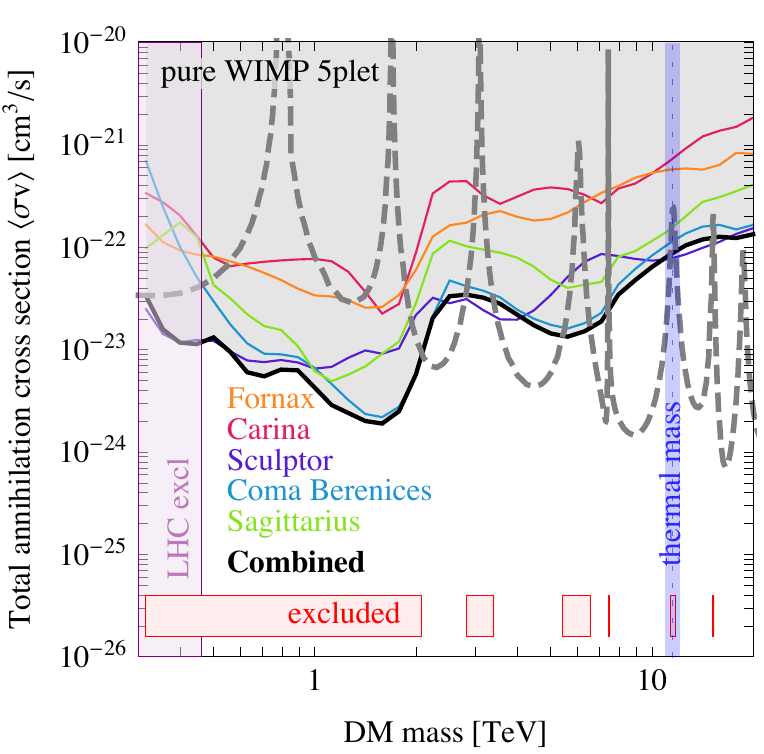} 
\caption{\label{fig:limitsmultiplet} 95\% C.L. exclusion limits on the total velocity-weighted annihilation cross section of the pure WIMP 3plet (left) and pure WIMP 5plet (right), as a function of the DM mass, using the full gamma-ray spectrum from these candidates. The predicted thermal mass of the candidate is indicated by a blue vertical band. The red boxes in the lower part of the plots indicate the areas in which the predicted cross section (gray dashed line) overshoots the combined limit (black thick line) and therefore the model is excluded. The shaded area at small DM masses is ruled out by disappearing track searches at the LHC.}
\end{figure}

The 95\% C.L. constraints on $\langle \sigma v \rangle$ for the two pure WIMP DM candidates are presented in Fig~\ref{fig:limitsmultiplet}.
As in Fig~\ref{fig:linelimits}, the limits from individual dSphs (colored lines) as well as from the combination (black line and shaded area) are shown. The total predicted annihilation cross section (the sum of the different contributions in Fig~\ref{fig:cross_sections}), to which the limit applies, is superimposed as a gray line. The regions where it overshoots the combined limit are excluded.  Also indicated with vertical bands is the mass predicted by thermal freeze-out.
The results are not limited to the freeze-out mass, both to have a quantification of the reach of this search, and because other production mechanisms (see e.g.~\cite{Moroi:1999zb}) or cosmological histories (see e.g.~\cite{Giudice:2000ex}) could result in
an EW multiplet with a different mass, while still constituting 100\% of the DM.

\smallskip

{ Finally, some  comments on the state of collider and direct detection searches for the pure-WIMP candidates considered in this study.
The LHC excludes values of DM mass smaller than 460~GeV~\cite{Aaboud:2017mpt}, via disappearing track searches.
The best limits to date~\cite{Aprile:2017iyp} on spin-independent DM-nucleon cross sections are not sensitive to any of the DM mass values that used in these studies, according to the predictions of~\cite{Hisano:2015rsa}.
}

\bigskip


\section{Discussion}
\label{sec:discussion}

\begin{figure}[!t]
\centering
\includegraphics[width=0.48\textwidth]{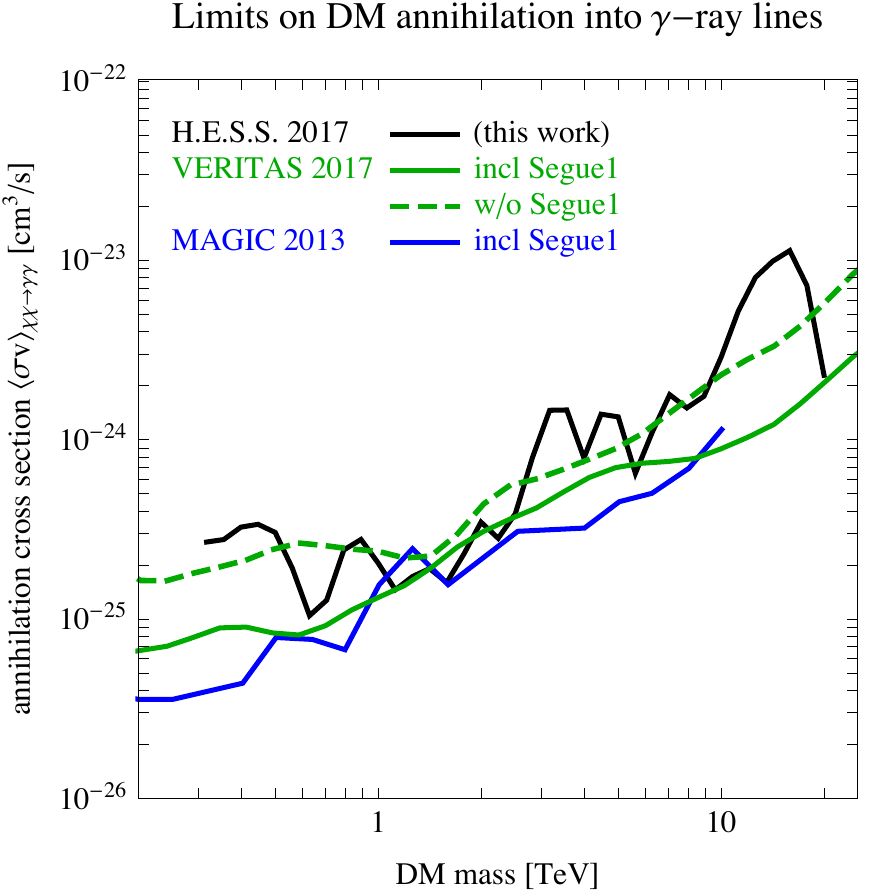} \quad
\includegraphics[width=0.48\textwidth]{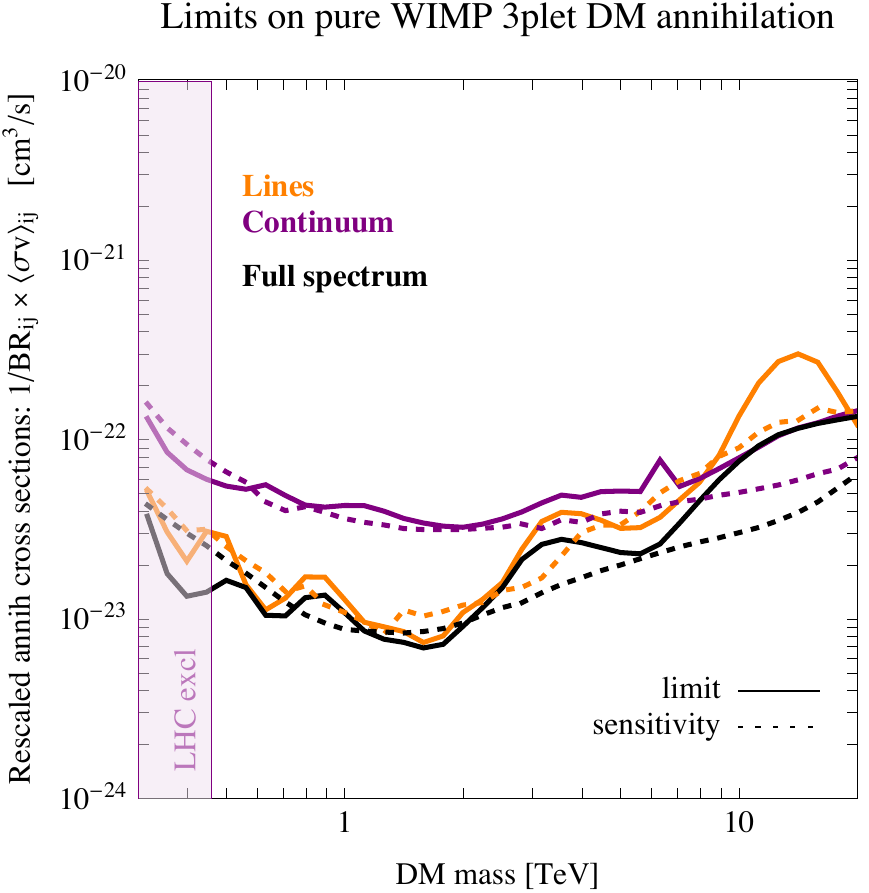}
\caption{\label{fig:comparison} {\em Left:} comparison of the limits on DM annihilation into mono-energetic gamma rays from different experiments. The H.E.S.S.~limits differ from those in Fig~\ref{fig:linelimits} since the $J$-factor nuisance is suppressed here for the purposes of this comparison. {\em Right:} limits on the annihilation cross section of the pure WIMP 3plet, computed in three different ways and rescaled by the corresponding branching ratios: the limit computed using only the line-like emission is rescaled by the BR($\gamma\gamma + \gamma Z/2$) (orange line); the limit computed using only the continuum gamma-ray emission from electroweak showering and gauge boson decays is rescaled by BR($WW + ZZ + \gamma Z/2$) (purple line); the limit using the full spectrum (black).}
\end{figure}

The constraints on DM annihilation into mono-energetic gamma rays (Fig.~\ref{fig:linelimits}) can be compared with the results by other collaborations, namely {\sc Magic}~\cite{Aleksic:2013xea}  { (with almost 160h of observation of Segue1)} and {\sc Veritas}~\cite{Archambault:2017wyh} { (with 230h of observation of 4 dSph's plus Segue1)}, keeping in mind that their limits are not obtained for the same source sample.
This is done in Fig.~\ref{fig:comparison} (left), where the H.E.S.S. constraints have been reprocessed in order to remove the error due to the uncertainty of the $J$-factor (see Sec.~\ref{sec:methodology}), as it is not included in the other analyses. Removing the nuisance factor overestimates the limits by a factor of $\sim$2. Without taking into account systematical uncertainties, the H.E.S.S. constraints are comparable to {\sc Magic} and {\sc Veritas} at round 1 TeV, and comparable to the {\sc Veritas} limits without Segue 1 on the whole mass range. Note however that the dSphs observed by H.E.S.S. do not include Segue 1, whose $J$-factor determination is subject to large uncertainties~\cite{Bonnivard:2015xpq} and might be much smaller than initially thought.  
\medskip

The gamma-ray line constraints can also be compared with the bounds obtained by H.E.S.S. {from} observations of the Galactic Center region~\cite{HESSlinesGC,Abdalla:2016olq,Abdallah:2018qtu}. The latter ones are much stronger, ranging around $\langle \sigma v \rangle \lesssim 10^{-(27 \ldots 28)} \ {\rm cm}^3/{\rm s}$ for $M_{\rm DM} \simeq$ 1 TeV, for a nominal Einasto DM profile.
However, as already commented above, they are much more dependent on the spatial shape of the DM profile in the inner Galaxy, for example they are relaxed by up to three orders of magnitude for cored profiles like the Burkert one (see e.g.~\cite{Cirelli:2015bda}). 

\medskip

A detailed comparison with the H.E.S.S. results on observation of continuum gamma rays from dSphs~\cite{Abramowski:2014tra,Abramowski:2010aa} has not been carried out, since the different annihilation channels are not systematically considered here and since that paper  employed a different set of analysis cuts and reconstruction technique. For the cases that can be compared (essentially the $WW$ channel) it is noted that the bounds derived in this work are comparable or slightly stronger than the ones in~\cite{Abramowski:2014tra}.
The improvements on the limits come mainly from the use of a more sensitive analysis chain and of spatial bins in the likelihood function of individual dSphs. The gain due to the additional spatial bins depend on the extension of the DM halo, and it reaches $\sim$10\% for Sculptor, Carina and Sagittarius, $\sim$20\% for Fornax, and $\sim$90\% for Coma Berenices limits.  

\medskip

Concerning the bounds on pure WIMPs this analysis constitutes, to the knowledge of the authors, the first where the full gamma-ray spectrum has been considered, as opposed to the usual searches for gamma-ray lines or continuum spectra (e.g.~from $WW$). It is therefore particularly interesting to ask how much the sensitivity is improved in this way. This is addressed in Fig.~\ref{fig:comparison} (right), where, for the case of the pure 3plet for definiteness, the limits on the annihilation cross sections in the three different cases (lines, continuum and full spectrum), rescaled by the corresponding branching ratios (respectively: BR($\gamma\gamma + \gamma Z/2$), BR($WW + ZZ + \gamma Z/2$ and 1), are shown. It can be seen that employing the full spectrum gives an improved sensitivity over the whole mass range.

\medskip

It is noted that the excluded region for the 3plet case is rather limited and does not directly affect the preferred mass of the candidate as determined by the thermal freeze-out process. Future observations, or future possible refinements of the theoretical computations of cross section and mass, have however the power of modifying the picture.
For the case of the 5plet, the excluded regions are more extended at low mass. At masses larger than $\sim$ 2 TeV they depend sensitively on the features in the annihilation cross section.
The thermal mass lies on a peak and therefore a portion of its uncertainty band is in the nominally excluded region. Here too, future possible refinements (notably concerning DM bound states formation) can modify the picture.

\medskip

The constraints obtained on pure WIMP candidates complement the results for this kind of models discussed in \cite{Cirelli:2014dsa,Cirelli:2015bda,Garcia-Cely:2015dda,Mitridate:2017izz} (and references therein). Generally speaking, the gamma-ray constraints from the GC region are more stringent, but also carry a much larger uncertainty related to the choice of the galactic DM profile. The recent limits from antiprotons~\cite{Cuoco:2017iax} are among the most stringent but are also subject to the analysis assumptions in~\cite{Cuoco:2016eej}, notably concerning the antiproton propagation and the treatment of secondary over primary CR ratios. If confirmed, they exclude large portions of the mass range for the 5plet and essentially the whole range for the case of the 3plet.

\medskip

The present analysis does not include any possible effect due to substructures or clumpiness in the observed target galaxies, since the determination of the $J$-factors adopted include the smooth halo only. While the existence of such clumps, and their specific properties, could induce a boost of the annihilation signal and lead to more stringent constraints, the inclusion of tidal stripping effects have been found to limit the boost to at most a few tens of percent (see e.g.~\cite{Moline:2016pbm}).



\section{Summary}
\label{sec:summary}

Dark Matter at the TeV and multi-TeV scale remains attractive as it is predicted in the right abundance by thermal freeze-out in the Early Universe. It remains therefore timely to pursue searches of DM annihilation in the corresponding energy regime. 
The data re-analyzed here were collected by the H.E.S.S. experiment from the observations of five {dSphs} (Fornax, Coma Berenices, Sculptor, Carina and Sagittarius), which have been targeted for a total time of about 130 hours. 
No significant excesses have been found, and therefore constraints are imposed on DM annihilation cross sections as a function of the DM mass, in the range 300~GeV - 20~TeV.
Under consideration were the model independent case of DM annihilating into mono-energetic gamma rays and the paradigmatic case of DM in the form of pure WIMP multiplets, annihilating into both $\gamma\gamma$ pairs and gauge bosons. The results for the gamma-ray lines are presented in Fig.~\ref{fig:linelimits}. A 95\%CL upper limit of $\langle \sigma v \rangle \lesssim 3 \times 10^{-25}$ cm$^3$ s$^{-1}$ is reached in the mass range of 400 GeV to 1 TeV. These bounds are compared to the results by other collaborations in Fig.~\ref{fig:comparison} (left). The results for the pure WIMP 3plet and 5plet are presented in Fig.~\ref{fig:limitsmultiplet}, with several excluded regions identified.
The analysis strategy of looking for the total spectrum of a model, improves over the usual searches for simple single channel final states ($\gamma\gamma$, $WW$) by a factor of up to a few tens of percent, as shown in Fig.~\ref{fig:comparison} (right). 
The same analysis strategy can be applied to future datasets and other observational regions.

\acknowledgments
\scriptsize
The support of the Namibian authorities and of the University of Namibia in facilitating 
the construction and operation of H.E.S.S. is gratefully acknowledged, as is the support 
by the German Ministry for Education and Research (BMBF), the Max Planck Society, the 
German Research Foundation (DFG), the Helmholtz Association, the Alexander von Humboldt Foundation, the French Ministry of Higher Education, Research and Innovation, the Centre National de la Recherche Scientifique (CNRS/IN2P3 and CNRS/INSU), the Commissariat à l'énergie atomique et aux énergies alternatives (CEA), the U.K. Science and Technology Facilities Council (STFC), the Knut and Alice Wallenberg Foundation, the National Science Centre, Poland grant no. 2016/22/M/ST9/00382, the South African Department of Science and Technology and National Research Foundation, the University of Namibia, the National Commission on Research, Science \& Technology of Namibia (NCRST), the Austrian Federal Ministry of Education, Science and Research and the Austrian Science Fund (FWF), the Australian Research Council (ARC), the Japan Society for the Promotion of Science and by the University of Amsterdam. We appreciate the excellent work of the technical support staff in Berlin, Zeuthen, Heidelberg, Palaiseau, Paris, Saclay, Tübingen and in Namibia in the construction and operation of the equipment. This work benefited from services provided by the H.E.S.S. Virtual Organisation, supported by the national resource providers of the EGI Federation. The work of A.V.~has been supported by the São Paulo Research Foundation (FAPESP) through Grant No 2015/15897-1, and it was financed in part by the Coordenação de Aperfeiçoamento de Pessoal de Nível Superior - Brasil (CAPES) - Finance Code 001. The work of M.C.~and F.S.~has been supported by the European Research Council ({\sc Erc}) under the EU Seventh Framework Programme (FP7/2007-2013) / {\sc Erc} Starting Grant (agreement n. 278234 - {\sc `NewDark'} project). The work of J.S.~has been supported in part by ERC Project No. 267117 ({\sc Dark}) hosted by the Pierre and Marie Curie University-Paris VI, Sorbonne University and CEA-Saclay and by the Institut Lagrange de Paris. The work of M.T.~is supported by the Spanish Research Agency (Agencia Estatal de Investigaci\'on) through the grant IFT Centro de Excelencia Severo Ochoa SEV-2016-0597 and the projects FPA2015-65929-P and Consolider MultiDark CSD2009-00064. This work has been done in part within the Labex ILP (reference ANR-10-LABX-63) part of the Idex SUPER, and received financial state aid managed by the Agence Nationale de la Recherche, as part of the programme {\it Investissements d'avenir} under the reference ANR-11- IDEX-0004-02. M.C.~and F.S.~acknowledge the hospitality of the Institut d'Astrophysique de Paris (IAP) where a part of this work was done. F.S. acknowledges partial support by a PIER Seed Project funding (Project ID PIF-2017-72).



\end{document}